\documentclass[12pt,a4paper,dvipdfmx]{amsart}
\usepackage{amsmath}
\usepackage{amsthm}
\usepackage{amsfonts}
\usepackage{amssymb}
\usepackage{latexsym}
\usepackage{amscd}
\usepackage{stmaryrd}
\usepackage{amsbsy}
\usepackage{ytableau}

\usepackage{pb-diagram}
\allowdisplaybreaks
\newcommand{\Vir}{\mathrm{Vir}}
\newcommand{\Z}{\mathbb{Z}}
\newcommand{\C}{\mathbb{C}}
\newcommand{\Lam}{|\Lambda\rangle}

\newtheorem{thm}{Theorem}[section]

\newtheorem{prp}[thm]{Proposition}
\newtheorem{conj}[thm]{Conjecture}

\theoremstyle{definition}
\newtheorem{dfn}[thm]{Definition}
\theoremstyle{remark}

\begin{document}

\title{Conformal blocks and Painlev\'e functions}
\author{Hajime Nagoya}
\address{Kanazawa University}
\email{nagoya@se.kanazawa-u.ac.jp}

\maketitle

\begin{abstract}      
This paper is based on my presentation at RIMS workshop on 
``Theory of Integrable Systems and Its Applications in Various Fields" held in Kyoto on 19--21, August 2015. 
The aim of the present paper is to give a short account of recent studies  on 
relations between conformal blocks in the two dimensional conformal field theory 
and Painlev\'e functions. In addition, we present a conjecture on a combinatorial expansion 
formula of the three-point irregular conformal block at 
an irregular singular point, with 
two regular singular points and one irregular singular point. Our 
conjectural expansion formula is written in terms of pairs of skew Young 
diagrams, while the four-point regular conformal block, by AGT correspondence, 
is written in terms of pairs of Young diagrams. 
\end{abstract}
\tableofcontents
\section{Introduction}

\subsection{Painlev\'e equations}
Aiming to obtain new special functions,  P. Painlev\'e classified 
second order nonlinear ordinary differential equations 
\begin{equation*}
R\left(t,y,\frac{dy}{dt},\frac{d^2y}{dt^2}\right)=0
\end{equation*}
whose movable singular points are pole only and obtained new six equations around 
the beginning of the twentieth century, which are now called 
Painlev\'e equations: 
\begin{align*}
\mathrm{P_I}\quad  y''=&6y^2+t,
\\
\mathrm{P_{II}}\quad y''=&2y^3+ty+\alpha,
\\
\mathrm{P_{III}}\quad  y''=&\frac{1}{y}y'^2-\frac{1}{y}y'+\frac{1}{t}(\alpha y^2+\beta)
+\gamma y^3+\frac{\delta}{y},
\\
\mathrm{P_{IV}}\quad  y''=&\frac{1}{2y}y'^2+\frac{3}{2}y^3+4ty^2+2(t^2-\alpha)y+\frac{\beta}{y},
\\
\mathrm{P_V}\quad y''=&\left(\frac{1}{2y}+\frac{1}{y-1}\right)y'^2-\frac{1}{t}y'+\frac{(y-1)^2}{t^2}
\left(\alpha y+\frac{\beta}{y}\right)
\\
&\displaystyle+\gamma\frac{y}{t}+\delta\frac{y(y+1)}{y-1},
\\
\mathrm{P_{VI}}\quad y''=&\frac{1}{2}\left(\frac{1}{y}+\frac{1}{y-1}+\frac{1}{y-t}\right)y'^2
-\left(\frac{1}{t}+\frac{1}{t-1}+\frac{1}{y-t}\right)y'
\\
 &\displaystyle +\frac{y(y-1)(y-t)}{t^2(t-1)^2}\left(\alpha+\beta\frac{t}{y^2}+\gamma\frac{t-1}{(y-1)^2}
+\delta\frac{t(t-1)}{(y-t)^2}\right). 
\end{align*}

Let us see a relation between Weierstrass $\wp$ function and $\mathrm{P_I}$. 
Replacing $t$ of $\mathrm{P_I}$ 
\begin{equation*}
\frac{d^2y}{dt^2}=6y^2+t 
\end{equation*}
with  $-g_2/2\in \C$,
 we obtain a second-order differential equation
\begin{equation}\label{eq-2nd-elliptic}
\frac{d^2y}{dt^2}=6y^2-\frac{1}{2}g_2, 
\end{equation}
which is derived by differentiating the differential equation 
\begin{equation*}
\left(\frac{dy}{dt}\right)^2=4y^3-g_2y-g_3. 
\end{equation*}
Hence, Weierstrass $\wp$ function is a solution to 
\eqref{eq-2nd-elliptic}. 
Weierstrass $\sigma$ function is defined by 
\begin{equation*}
\wp=-\frac{d^2}{dt^2}\log \sigma. 
\end{equation*}
The first Painlev\'e equation is 
\begin{equation*}
\frac{d^2y}{dt^2}=6y^2+t. 
\end{equation*}
For any solution $\lambda(t)$, define $\tau(t)$ by 
\begin{equation*}
\lambda(t)=-\frac{d^2}{dt^2}\log \tau(t). 
\end{equation*}
For the other Painlev\'e equations, the tau functions are defined in the same way. 
As theta functions play an important role in the theory of elliptic functions, 
the tau functions $\tau(t)$ are  relevant tools for the studies 
of the Painlev\'e functions. We remark that 
$\mathrm{P_I}$ is a Hamiltonian system with 
\begin{equation*}
H(t)=\frac{1}{2}\mu(t)^2-2\lambda(t)^3-t\lambda(t)\quad 
\left(\mu(t)=\frac{d\lambda(t)}{dt}\right). 
\end{equation*}
Note that $H(t)=(\log \tau(t))'$. 
$H(t)$ satisfies a non-linear differential equation
\begin{equation*}
\frac{d^3H}{dt^3}+6\left(\frac{dH}{dt}\right)^2+t=0. 
\end{equation*}
Conversely, a function $H(t)$ satisfying the differential equation above 
defines a solution $\lambda(t)$ to the first Painlev\'e equation.

\subsection{Fourier expansion}

For a long time, it had not been known an explicit series representation 
of the tau functions of Painlev\'e equations, until a remarkable 
discovery by Gamayun, Iorgov and Lisovyy \cite{GIL1}. They found 
that the tau function of the sixth Painlev\'e equation 
is a Fourier expansion in terms of the four point Virasoro conformal block 
with central charge $c=1$. 
Conformal blocks are the building blocks of correlation functions of the two dimensional 
conformal field theory, which are defined as 
expectation values of vertex operators. 
 
Their formula reads as
\begin{equation}\label{PVI regular}
\tau_{\mathrm{VI}}(t)=\sum_{n\in\mathbb{Z}}
s^n C\left(\begin{matrix}
\theta_1, \theta_t
\\
\theta_\infty,\sigma+n,\theta_0
\end{matrix}\right)\mathcal{F}\left(\begin{matrix}
\theta_1, \theta_t
\\
\theta_\infty,\sigma+n,\theta_0\end{matrix};t\right), 
\end{equation}
where $s,\sigma\in\mathbb{C}$, $\mathcal{F}(\theta,\sigma;t)=t^{\sigma^2-\theta_t^2-\theta_0^2}(1+O(t))$ is the four-point Virasoro conformal block 
with $c=1$,  and 
\begin{equation*}
C(\theta, \sigma)=\frac{\prod_{\epsilon,\epsilon'=\pm}
G(1+\theta_t+\epsilon \theta_0+\epsilon' \sigma)
G(1+\theta_1+\epsilon \theta_\infty
+\epsilon' \sigma)}{\prod_{\epsilon=\pm}G(1+2 \epsilon \sigma)},
\end{equation*} 
where $G(z)$ is the Barnes G-function such that $G(z+1)=\Gamma(z)G(z)$. 
By AGT correspondence, the four-point Virasoro conformal block 
has an explicit series representation \eqref{eq_AGT_CB}. 

Before this discovery, it was known by Jimbo \cite{Jimbo 1982} that 
 the tau function of the sixth Painlev\'e equation is expanded as 
\begin{align*}
\tau(t)=&\ \mathrm{const.}\ t^{(\sigma^2-\theta_0^2-\theta_t^2)}
\\
\times &\left( 1+\frac{(\theta_0^2-\theta_t^2-\sigma^2)
(\theta_\infty^2-\theta_1^2-\sigma^2)}{2 \sigma^2}t\right.
\\
&-\sum_{\epsilon=\pm}\frac{\hat{s}^\epsilon}{8\sigma^2(1+4\sigma^2)^2}
(\theta_0^2-(\theta_t-\epsilon\sigma)^2) 
(\theta_\infty^2-(\theta_1-\epsilon\sigma)^2)t^{1+2\epsilon \sigma}
\\
&\left. +\sum_{j=2}^\infty \sum_{|k|\le j}a_{jk}t^{j-2k\sigma}\right), 
\end{align*}
where $\hat{s}$, $\sigma$ are expressed by  
the monodromy data of a linear equation associated 
with $\mathrm{P_{VI}}$. The first part of $\tau(t)$:  
\begin{equation*}
t^{\sigma^2-\theta_0^2-\theta_t^2}
\left( 1+\frac{(\theta_0^2-\theta_t^2-\sigma^2)
(\theta_\infty^2-\theta_1^2-\sigma^2)}{2 \sigma^2}t\right)
\end{equation*}
resembles the first part of the four-point conformal block of Virasoro CFT:  
\begin{align*}
&\langle \Delta_4|\cdot\left( \Phi^{\Delta_3}_{\Delta_4,\Delta}(1)
\circ\Phi^{\Delta_2}_{\Delta,\Delta_1}(t)
|\Delta_1\rangle \right)
\\
&=t^{\Delta-\Delta_2-\Delta_1}
\left(1+\frac{(\Delta+\Delta_3-\Delta_4)(\Delta+\Delta_2-\Delta_1)}{2\Delta}t
+\sum_{k=2}^\infty c_k\left(\frac{z_2}{z_3}\right)^k
\right). 
\end{align*}
The definitions of a vertex operator and a conformal block 
are given in Section 2. Note that when $c=1$, parametrizing $\Delta_i=\theta_i^2$ 
is standard. 
It seems that this had been missed for a long time. Finally,  
it was noticed and leaded them to the discovery. 

Proofs for the Fourier expansion of the tau function of $\mathrm{P_{VI}}$ 
in terms of the four-point conformal block were 
given in
\cite{Iorgov Lisovyy Teschner} by constructing 
a fundamental solution of a linear equation associated 
with $\mathrm{P_{VI}}$ using the conformal field theory, 
and 
\cite{Bershtein Shchechkin} by showing that the Fourier expansion 
satisfies the bilinear equation for $\tau_{\mathrm{VI}}$ 
using embedding the direct sum of two Virasoro algebras into 
the super Virasoro algebra. 

\subsection{Irregular case}

It is well known that Painlev\'e equations admit 
the following degeneration scheme:
\begin{equation*}
\begin{diagram}
\node{\mathrm{P_{VI}}}\arrow{e}
\node{\mathrm{P_{V}}}\arrow{e}\arrow{se}
\node{\mathrm{P_{III}}}\arrow{e}\arrow{se}
\node{\mathrm{P_{III}^{D_7}}}\arrow{e}\arrow{se}
\node{\mathrm{P_{III}^{D_8}}}
\\
\node[3]{\mathrm{P_{IV}}}\arrow{e}
\node{\mathrm{P_{II}}}\arrow{e}
\node{\mathrm{P_I}}
\end{diagram}
\end{equation*}
Looking at the Painlev\'e equations, we know the fixed singular 
points are as in Table 1. 
\begin{table}[htb]
\begin{center}
\caption{}
\begin{tabular}{|l|c|}
\hline
& singular points
\\ \hline
$\mathrm{P_{VI}}$& $0,1,\infty$
\\ \hline
$\mathrm{P_{V}}$, $\mathrm{P_{III}}$, $\mathrm{P_{III}^{D_7}}$, 
$\mathrm{P_{III}^{D_8}}$ & $0,\infty$
\\ \hline
$\mathrm{P_{IV}}$, $\mathrm{P_{II}}$, $\mathrm{P_{I}}$& $\infty$
\\ \hline
\end{tabular}
\end{center}
\end{table}
Here, $\infty$ is an irregular singular point for Painlev\'e equations 
except $\mathrm{P_{VI}}$, and $0$ is a regular singular 
point for $\mathrm{P_{VI}}$, $\mathrm{P_{V}}$, $\mathrm{P_{III}}$, $\mathrm{P_{III}^{D_7}}$ and  
$\mathrm{P_{III}^{D_8}}$. In general, it is easier to compute 
confluent process of a series expansion at a regular singular point 
than at an irregular singular point. This is true for our case. 
Series expansions at the regular singular point $0$ of the tau functions 
of $\mathrm{P_{V}}$, $\mathrm{P_{III}}$, $\mathrm{P_{III}^{D_7}}$ and  
$\mathrm{P_{III}^{D_8}}$ 
were obtained in \cite{GIL2} by taking some collision limits. 
It has not been reported at the moment 
that  expansions at the irregular singular point $\infty$ 
of the tau functions of Painlev\'e equations 
are obtained by some confluent process. Instead, 
it was conjectured that a Fourier expansion at $\infty$ 
of the tau function for $\mathrm{P_{III}^{D_8}}$ 
can be computed recursively and the first few terms were obtained  
in \cite{Its Lisovyy Tykhyy 2014}. 

It is natural to expect that series expansions at $\infty$ 
of the tau functions are Fourier expansions in terms of irregular conformal blocks. 
Since we lacked enough knowledge on irregular conformal blocks, 
Fourier expansions of the tau functions in terms of irregular conformal blocks 
were not given. 
We need to understand 
expansions of irregular conformal blocks at irregular singular points 
more clearly. From this point of view, a precise definition of irregular vertex operators 
were introduced and irregular conformal blocks were 
given as expectation values of irregular vertex operators \cite{N ICB}. 
Using newly introduced irregular conformal blocks, the author was able to 
obtain conjectural formulas for Fourier expansions at $\infty$ 
of the tau functions of $\mathrm{P_{V}}$ and $\mathrm{P_{IV}}$ \cite{N ICB}. 

The remainder of this paper is organized as follows. 
In Section 2, a short review on regular conformal blocks are given. 
In Section 3, we recall what was known about 
irregular conformal blocks before \cite{N ICB}. 
In Section 4, a direct approach for obtaining irregular conformal blocks 
are explained, following \cite{N ICB}. 
In Section 5, we present conjectural formulas for Fourier expansions at $\infty$ 
of the tau functions of $\mathrm{P_{V}}$ and $\mathrm{P_{IV}}$ 
proposed in \cite{N ICB}  and give comments on a combinatorial formula  
for three-point irregular conformal block with 
two regular singular points and one irregular singular point of rank $1$.

\section{Regular conformal blocks}

The Virasoro algebra 
\begin{equation*}
\Vir=\bigoplus_{n\in\Z}\C L_n\oplus \C C
\end{equation*}
is the Lie algebra with commutation relations:
\begin{align*}
&[L_m, L_n]=(m-n)L_{m+n}+\frac{1}{12}(m^3-m)\delta_{m+n,0}C,
\\
&[\Vir, C]=0,
\end{align*}
where $\delta_{i,j}$ stands for Kronecker's delta. 
We use the following vertex operator to describe the behavior of the conformal block at a regular singular point.  
A Verma module $V_{\Delta}$ with the highest weight  $\Delta$ of the Virasoro algebra   is the induced module 
\begin{equation*}
V_{\Delta}=\mathrm{Ind}_{\Vir_{\ge 0}}^\Vir \C \Lam\ (=U(\Vir)\otimes_{\Vir_{\ge 0}}\C \Lam). 
\end{equation*}
Denote the dual Verma module by $V^*_\Delta$ such that  
\begin{align*}
&\langle \Delta | L_0=\Delta\langle \Delta |, 
\quad \langle \Delta | L_{-n}=0\quad (n>0). 
\end{align*} 
A bilinear pairing $\langle |\rangle$:  $V^*_\Delta\times V_\Delta\to \C$ is uniquely defined by 
\begin{align*}
&\langle \Delta| \cdot |\Delta\rangle=1,
\\
&\langle u|L_n \cdot |v\rangle =\langle u| \cdot L_n |v\rangle \equiv 
\langle u| L_n |v\rangle\quad (n\in\Z), 
\end{align*}
where $u\in V^*_\Delta$ and $v\in V_\Delta$. 

\begin{dfn}
Let an operator $\Phi_{\Delta_3,\Delta_1}^{\Delta_2}(z)$: $V_{\Delta_1}\to V_{\Delta_3}$ be defined by 
\begin{align}
&[L_n, \Phi_{\Delta_3,\Delta_1}^{\Delta_2}(z)]=z^n \left(z\frac{\partial}{\partial z}+
(n+1)\Delta_2\right)\Phi_{\Delta_3,\Delta_1}^{\Delta_2}(z), \label{comrel_rank0}
\\
&\Phi_{\Delta_3,\Delta_1}^{\Delta_2}(z)|\Delta_1\rangle=
z^{\Delta_3-\Delta_2-\Delta_1} 
\sum_{m\in\Z_{\ge 0}} v_mz^m,\label{eq_VO_act}
\end{align}
where  $v_m\in V_{\Delta_3}$ and $v_0=|\Delta_3\rangle$. 
\end{dfn}
We call $\Phi_{\Delta_3,\Delta_1}^{\Delta_2}(z)$ a vertex operator. 
For $u\in V_\Delta$, since 
\begin{align*}
&\Phi_{\Delta_3,\Delta_1}^{\Delta_2}(z)L_{n}u
\\
&=[\Phi_{\Delta_3,\Delta_1}^{\Delta_2}(z),L_{n}]u+L_{n}\Phi_{\Delta_3,\Delta_1}^{\Delta_2}(z)u
\\
&=-z^n\left(z\frac{\partial}{\partial z}+(n+1)\Delta_2\right)\Phi_{\Delta_3,\Delta_1}^{\Delta_2}(z)u
+L_{n}\Phi_{\Delta_3,\Delta_1}^{\Delta_2}(z)u,  
\end{align*}
we only need to determine the action of 
the vertex operator $\Phi_{\Delta_3,\Delta_1}^{\Delta_2}(z)$ on the highest weight $|\Delta_1\rangle$, 
namely,  $v_m$ for $m\ge 1$.  
The relation \eqref{comrel_rank0} 
on $|\Delta_1\rangle$ for $n\ge 0$ is equal to 
\begin{equation}\label{eq_action_L_n}
L_nv_m=(\Delta_3+n\Delta_2-\Delta_1+m-n+\delta_{n,0}\Delta_1)v_{m-n}.
\end{equation}
Considering the case of $n=0$, we see 
$v_m\in V_m=\{v\in V_{\Delta_3}|L_0v=(\Delta_3+m)v\}$ ($m\ge 0$). A basis of the eigenspace $V_m$ of $L_0$  
is $\{L_{-\lambda}|\Delta_3\rangle||\lambda|=m\}$, where 
$\lambda=(\lambda_1,\ldots,\lambda_k)$  
($\lambda_1\ge \lambda_2\ge \cdots \ge \lambda_k>0$) 
is a partition of a positive integer, 
$|\lambda|= \sum_{i=1}^k\lambda_i$ and 
$L_{-\lambda}=L_{-\lambda_1}\cdots L_{-\lambda_k}$. 
Set $v_m=\sum_{\lambda,|\lambda|=m}c_\lambda 
L_{-\lambda}|\Delta_3\rangle$. We have 
\begin{equation*}
\left(\langle \Delta_3 |L_\lambda v_m\right)_{|\lambda|=m}
=\left(\langle \Delta_3|L_\lambda
L_{-\mu}|\Delta_3\rangle\right)_{ |\lambda|=|\mu|=m} 
\cdot \left(c_\mu\right)_{|\mu|=m},
\end{equation*}
where $L_\lambda=L_{\lambda_k}\cdots L_{\lambda_1}$. 
Hence, if the Kac determinant $\left|\langle \Delta_3|L_\lambda
L_{-\mu}|\Delta_3\rangle\right|_{ |\lambda|=|\mu|=m}$ 
is not zero, then the coefficients $c_\mu$ ($|\mu|=m$)  
are uniquely determined recursively by \eqref{eq_action_L_n}. It follows from the 
construction that $v_m$ ($m\ge 0$) satisfy 
the relation \eqref{eq_action_L_n}. 
Since the condition that all Kac determinants are not zero, 
is equivalent to irreducibility of a Verma module, 
we have the following proposition.  
\begin{prp}
If the Verma module $V_{\Delta_3}$ is irreducible,  then 
the vertex operator $\Phi_{\Delta_3,\Delta_1}^{\Delta_2}(z)$ 
exists uniquely. 
\end{prp}
We define the $n$-point conformal block  as the expectation value of the vertex operators:
\begin{equation*}
\langle \Delta_n|\cdot \left(
\Phi^{\Delta_{n-1}}_{\Delta_{n},\widetilde{\Delta}_{n-3}}(z_{n-1})
\circ\cdots\circ
\Phi^{\Delta_2}_{\widetilde{\Delta}_1,\Delta_1}(z_2)
|\Delta_1\rangle\right),  
\end{equation*} 
where we suppose the Verma modules $V_{\widetilde{\Delta}_i}$ ($i=1,\ldots,n-2$) and $V_{\Delta_n}$ are 
irreducible. These conformal blocks were 
introduced in \cite{BPZ}, are formal power 
series in $z_i/z_{i+1}$ ($i=2,\ldots,n-2$) and are believed to be absolutely convergent in the domain 
$|z_2|<\cdots<|z_{n-1}|$. As an example, let us 
see the first part of the four-point conformal block 
\begin{align*}
&\langle \Delta_4|\cdot\left( \Phi^{\Delta_3}_{\Delta_4,\Delta}(z_3)
\circ\Phi^{\Delta_2}_{\Delta,\Delta_1}(z_2)
|\Delta_1\rangle \right)
\\
&=z_3^{\Delta_4-\Delta_3-\Delta_2}z_2^{\Delta-\Delta_2-\Delta_1}
\left(1+\frac{(\Delta+\Delta_3-\Delta_4)(\Delta+\Delta_2-\Delta_1)}{2\Delta}\frac{z_2}{z_3}
+\sum_{k=2}^\infty c_k\left(\frac{z_2}{z_3}\right)^k
\right). 
\end{align*}
The first term $c_1$ is factorized but the second term $c_2$ is already complicated, and we had not 
known an explicit formula for coefficients $c_k$ until recently. 
In 2009, Alday, Gaiotto and Tachikawa conjectured a correspondence between Virasoro conformal blocks 
of 2d Liouville CFT and the Nekrasov partition function of a certain class of 4d supersymmetric 
gauge theories \cite{AGT}. 
This correspondence was proved by Alba, Fateev, Litvinov and Tarnopolskiy
\cite{AFLT}.  
We present the explicit series representation of 
the four-point conformal block for the central charge $c=1$. 
Put 
\begin{equation*}
\Delta_1=\theta_0^2,\quad \Delta_2=\theta_t^2,\quad 
\Delta=\sigma^2,\quad \Delta_3=\theta_1^2,\quad 
\Delta_4=\theta_\infty^2,\quad z_2=t,\quad z_3=1. 
\end{equation*}
Then, by AGT correspondence,  
we have 
\begin{equation}\label{eq_AGT_CB}
\langle \theta_\infty^2|\cdot\left( \Phi^{\theta_1^2}_{\theta_\infty^2,\sigma^2}(1)
\circ\Phi^{\theta_t^2}_{\sigma^2,\theta_0^2}(t)
|\theta_0^2\rangle \right)
=
t^{\sigma^2-\theta_0^2-\theta_t^2}(1-t)^{2\theta_0\theta_1}
\sum_{\lambda, \mu\in \mathbb{Y}}N_{\lambda,\mu}
\left(\begin{matrix}
\theta_1, \theta_t
\\
\theta_\infty,\sigma,\theta_0\end{matrix}\right)t^{|\lambda|+|\mu|},
\end{equation}
where $\mathbb{Y}$ stands for the set of all Young diagrams, 
\begin{align*}
N_{\lambda,\mu}
\left(\begin{matrix}
\theta_1, \theta_t
\\
\theta_\infty,\sigma,\theta_0\end{matrix}
\right)
=&\prod_{(i,j)\in\lambda}
\frac{((\theta_t+\sigma+i-j)^2-\theta_0^2)((\theta_1+\sigma+i-j)^2-\theta_\infty^2)}
{h_\lambda^2(i,j)(\lambda_j'+\mu_i-i-j+1+2\sigma)^2}
\\
&\times\prod_{(i,j)\in\mu}
\frac{((\theta_t-\sigma+i-j)^2-\theta_0^2)((\theta_1-\sigma+i-j)^2-\theta_\infty^2)}
{h_\mu^2(i,j)(\mu_j'+\lambda_i-i-j+1-2\sigma)^2}, 
\end{align*}
and $\lambda=(\lambda_1,\ldots,\lambda_n)$ ($\lambda_1\ge \lambda_2\ge\cdots\ge \lambda_n>0$), $|\lambda|=\sum_{i=1}^n\lambda_i$,   
$\lambda'$ denotes the transposition of $\lambda$, and $h_\lambda(i,j)$ is the hook length 
defined by $h_\lambda(i,j)=\lambda_i+\lambda_j'-i-j+1$. 
We note that the above series expression is a generalization of 
Gauss's hypergeometric series. If we set $\theta_t=1/2$, 
then $\mu$ must be the empty set $\{\phi\}$ and if $\sigma=\theta_0+1/2$, 
then $\lambda$ must be $(1^n)$ ($n\in\mathbb{Z}_{\ge 0}$). Hence, 
when $\theta_t=1/2$ and $\sigma=\theta_0+1/2$, we have 
\begin{equation*}
\sum_{\lambda, \mu\in \mathbb{Y}}N_{\lambda,\mu}
\left(\begin{matrix}
\theta_1, 1/2
\\
\theta_\infty,\theta_0+1/2,\theta_0\end{matrix}
\right)t^{|\lambda|+|\mu|}
=\sum_{n=0}^\infty t^n
\prod_{i=1}^n \frac{((\theta_1+\theta_0+i-1/2)^2-\theta_\infty^2)}{n!(2\theta_0+i)}. 
\end{equation*}

\section{Known results of irregular conformal blocks}

\subsection{Integral representation}

The Knizhnik-Zamolodchikov (KZ) equations 
satisfied by the correlation functions 
of the conformal field theory associated with 
 an affine Lie algebra, are quantization 
 of the Schlesinger equations \cite{R}, \cite{H}. 
 For $\widehat{\mathfrak{sl}}_2$ case, solutions 
 to KZ equations are corresponding to 
 Virasoro conformal blocks \cite{RT}, \cite{N HGS}. 
The  Schlesinger equation describes 
isomonodromic deformation of a Fuchsian system 
of regular singular type. Quantizations of 
irregular versions 
of the Schlesinger equations 
were performed in \cite{FMTV}, \cite{BK} for
Poincar\'e rank one case and in \cite{JNS} 
for any Poincar\'e rank case associated with 
 $\mathfrak{sl}_2$. Solutions to 
 irregular versions of KZ equations above were given 
 by integral representations of 
 confluent hypergeometric type. 
 
 In \cite{Nagoya Sun 2010},  
 such integral representations were constructed
 by free field realizations of the confluent 
 primary fields.  
Since this method is applicable to  the 
 Virasoro case, we have integral formulas for irregular (Virasoro) conformal blocks. These 
 are generalizations of integral formulas 
 for regular conformal blocks 
  obtained  
 by Dotsenko and Fateev \cite{DF}. 
 
 We remark that integral formulas (of Dotsenko-Fateev type) exist for 
 particular conformal blocks only. As an example, 
 for the four-point conformal block, integral 
 formulas exist if and only if two parameters within five parameters 
 are parametrized by non-negative integers, as we did it to obtain 
 Gauss's hypergeometric series in the previous subsection. 
 Hence, we know global representations of 
 irregular conformal blocks for Virasoro algebra 
 in special cases. 
 
 As an example, an integral formula of a three-point function with 
 two regular singular points and one irregular singular point of rank one is 
 of the form 
 \begin{equation*}
 \int_0^1\cdots\int_0^1 dt_1\cdots dt_n
 \prod_{1\le i<j\le n}(t_i-t_j)^{2\beta}\prod_{i=1}^n
 t_i^u (1-t_i)^v e^{zt_i}, 
 \end{equation*}
 which is a generalization of the integral representation of 
 Kummer's confluent hypergeometric function. 
 
\subsection{Pairings of irregular vectors embedded in Verma modules} 
 
 For general case, the following method was introduced in \cite{Gaiotto}
 and developed in \cite{BMT}, \cite{GT}. 
It is to  take a pairing of 
irregular vectors $|\Lambda\rangle$ and $\langle \Lambda' |$ 
embedded in a Verma module and a dual Verma module, respectively, 
such that for non-negative integers $r,s,$ and tuples $\Lambda=(\Lambda_r,\Lambda_{r+1},\ldots, 
\Lambda_{2r})$, $\Lambda'=(\Lambda'_s,\Lambda'_{s+1},\ldots, 
\Lambda'_{2s})$, 
\begin{align}
&L_n |\Lambda\rangle=\Lambda_n |\Lambda\rangle \quad (n=r,\ldots, 2r),
\quad L_n |\Lambda\rangle=0\quad (n>2r), \label{eq_irregular_vector}
\\
&\langle \Lambda' |L_n=\Lambda'_n\langle \Lambda' |\quad (n=-s,\ldots, -2s),
\quad \langle \Lambda' |L_n=0 \quad (n<-2s).  \nonumber
\end{align}  
For $n=0$, the identity \eqref{eq_irregular_vector} is a well-known condition 
of the highest weight vector of the Verma module and is equal to 
the action of $L_n$ on the primary field. 
For $n>0$, the identities \eqref{eq_irregular_vector}  are equal to the actions of $L_n$ 
on the confluent primary field of rank $r$  ((3.1) in \cite{Nagoya Sun 2010}). It is 
natural to think that \eqref{eq_irregular_vector} is a condition 
of an irregular version of the highest weight vector. Hence, 
we expect that a pairing of such vectors produces an irregular conformal block. 

When $r=1$, the irregular vector $|\Lambda\rangle$ 
in a Verma module of the Virasoro algebra was obtained by 
a degeneration limit of the image $\Phi_{\Delta_3,\Delta_1}^{\Delta_2}(z)|\Delta_1\rangle$ 
of a highest weight vector by a vertex operator 
 in \cite{MMM} and given in terms of Jack symmetric functions 
\cite{Yanagida}. When $r>1$, 
the construction of irregular vectors was performed in \cite{FJK} and it was 
 revealed that the first-order irregular vectors are uniquely determined 
by the condition \eqref{eq_irregular_vector}, up to a scalar.  However, the higher-order irregular vectors contain 
an infinite number of parameters, and thus they are not unique. Consequently, 
irregular conformal blocks as pairings of irregular vectors are not 
uniquely determined by \eqref{eq_irregular_vector}. 
 In \cite{GT}, 
the authors proposed a method of constructing irregular vectors by delicate
 limiting procedures from the image of  a highest weight vector 
 by compositions of vertex operators. Hence, if their method 
 is justified, then we have uniquely determined irregular conformal blocks. 

In what follows, we demonstrate how to take a limit of 
$\Phi_{\Delta_3,\Delta_1}^{\Delta_2}(z)|\Delta_1\rangle$  as 
$z$ goes to zero, 
following \cite{MMM}.  

From the case of $n=1,2$ of \eqref{eq_action_L_n}: 
\begin{align}
&L_1v_m=(\Delta_3+\Delta_2-\Delta_1+m-1)v_{m-1},\label{eq_L_1}
\\
&L_2v_m=(\Delta_3+2\Delta_2-\Delta_1+m-2)v_{m-2}, \label{eq_L_2}
\end{align}
if we set 
\begin{align}
\Delta_2-\Delta_1=&c_1\Lambda+c_{10}+O(\Lambda^{-1}),
\label{eq_L_1-L}
\\
2\Delta_2-\Delta_1=&c_2\Lambda^2+c_{21}\Lambda+c_{20}+O(\Lambda^{-1}),
\label{eq_L_2-L}
\end{align}
we obtain $v_m=O(\Lambda^m)$ ($\Lambda\to \infty$). 
Hence, when $z$ goes to zero as $z=w/\Lambda$ ($\Lambda\to \infty$), 
$v_m z^m$ becomes finite. The relations 
\eqref{eq_L_1} and \eqref{eq_L_2} 
transform to
\begin{equation}\label{eq_L_1_L_2}
L_1p_m=c_1p_{m-1},\quad L_2p_m=c_2p_{m-2}, 
\end{equation}
where $v_m/\Lambda^m\to p_m$ ($\Lambda\to \infty$). 
Put $|W\rangle=\sum_{m=0}^\infty p_mw^m$. Then, we have 
\begin{equation*}
L_1|W\rangle=c_1w|W\rangle,\quad L_2|W\rangle=c_2w^2|W\rangle
\end{equation*}
from \eqref{eq_L_1_L_2}. 

We remark that when the central charge $c=1$, 
the conditions \eqref{eq_L_1-L} and \eqref{eq_L_2-L} 
reduce to \eqref{0000to100} if $c_2\neq 0$ and to 
\eqref{0000to100} plus \eqref{100to(1/2)00} if $c_2=0$.

\subsection{Degeneration limits of the Nekrasov partition functions}

On the gauge theory side, it is easy to compute degeneration limits of 
the Nekrasov partition functions. The process of the limit is called   decoupling   of   the   matter
 hypermultiplets in the language of the gauge theory. 
Let us see that by how to take successive limits  
of $N_{\lambda,\mu}$. 

Looking at factors $\theta_t\pm\sigma-\theta_0$ and $\theta_t\pm\sigma+\theta_0$ in $N_{\lambda,\mu}$, we find that 
\begin{equation}\label{0000to100}
\theta_1+\theta_\infty=\Lambda,\quad \theta_1-\theta_\infty=\theta_*,\quad t\to\frac{t}{\Lambda},\quad \Lambda\to \infty
\end{equation}
 transforms $N_{\lambda,\mu}t^{|\lambda|+|\mu|}$ to  
\begin{align*}
N_{\lambda,\mu}
\left(
\theta_0,
\theta_t,\sigma,\theta_*\right)t^{|\lambda|+|\mu|}
=&t^{|\lambda|+|\mu|}\prod_{(i,j)\in\lambda}
\frac{(\theta_*+\sigma+i-j)((\theta_t+\sigma+i-j)^2-\theta_0^2)}
{h_\lambda^2(i,j)(\lambda_j'+\mu_i-i-j+1+2\sigma)^2}
\\
&\times\prod_{(i,j)\in\mu}
\frac{(\theta_*-\sigma+i-j)((\theta_t-\sigma+i-j)^2-\theta_0^2)}
{h_\mu^2(i,j)(\mu_j'+\lambda_i-i-j+1-2\sigma)^2}. 
\end{align*}
 The degenerate function 
$\sum_{\lambda, \mu\in \mathbb{Y}}N_{\lambda,\mu}
\left(
\theta_0,
\theta_t,\sigma,\theta_*\right)t^{|\lambda|+|\mu|}$ 
is the builiding block of the tau function of the fifth Painlev\'e function at $t=0$ \cite{GIL2}. 

Next, we can take two limits. One is the same limit  
\begin{equation}\label{100to11}
\theta_t+\theta_0=\Lambda,\quad \theta_t-\theta_0=\theta_\star,
\quad t\to\frac{t}{\Lambda},\quad \Lambda\to\infty
\end{equation}
as above. We get from $N_{\lambda,\mu}
\left(
\theta_0,
\theta_t,\sigma,\theta_*\right)t^{|\lambda|+|\mu|}$
\begin{align*}
N_{\lambda,\mu}
\left(
\theta_\star,\sigma,\theta_*\right)t^{|\lambda|+|\mu|}
=&t^{|\lambda|+|\mu|}\prod_{(i,j)\in\lambda}
\frac{(\theta_*+\sigma+i-j)(\theta_\star+\sigma+i-j)}
{h_\lambda^2(i,j)(\lambda_j'+\mu_i-i-j+1+2\sigma)^2}
\\
&\times\prod_{(i,j)\in\mu}
\frac{(\theta_*-\sigma+i-j)(\theta_\star-\sigma+i-j)}
{h_\mu^2(i,j)(\mu_j'+\lambda_i-i-j+1-2\sigma)^2}.  
\end{align*}
 The degenerate function 
$\sum_{\lambda, \mu\in \mathbb{Y}}N_{\lambda,\mu}
\left(
\theta_\star,\sigma,\theta_*\right)t^{|\lambda|+|\mu|}$ 
is the builiding block of the tau function of the third Painlev\'e function at $t=0$ \cite{GIL2}. 
Another one is that 
\begin{equation}\label{100to(1/2)00}
\theta_*=\Lambda,\quad t\to \frac{t}{\Lambda},\quad \Lambda\to\infty.  
\end{equation}
 We get from $N_{\lambda,\mu}
\left(
\theta_0,
\theta_t,\sigma,\theta_*\right)t^{|\lambda|+|\mu|}$
\begin{align*}
N_{\lambda,\mu}
\left(
\theta_0,
\theta_t,\sigma\right)t^{|\lambda|+|\mu|}
=&t^{|\lambda|+|\mu|}\prod_{(i,j)\in\lambda}
\frac{((\theta_t+\sigma+i-j)^2-\theta_0^2)}
{h_\lambda^2(i,j)(\lambda_j'+\mu_i-i-j+1+2\sigma)^2}
\\
&\times\prod_{(i,j)\in\mu}
\frac{((\theta_t-\sigma+i-j)^2-\theta_0^2)}
{h_\mu^2(i,j)(\mu_j'+\lambda_i-i-j+1-2\sigma)^2}. 
\end{align*}
 If we set $\theta_t+\theta_0=\theta_*$ 
 and $\theta_t-\theta_0=\theta_\star$, then 
 $N_{\lambda,\mu}
\left(
\theta_0,
\theta_t,\sigma\right)$ coincides with 
$N_{\lambda,\mu}
\left(
\theta_\star,\sigma,\theta_*\right)$. 

Next, by the limit
\begin{equation}\label{11to(1/2)1}
\theta_*=\Lambda,\quad t\to \frac{t}{\Lambda},\quad \Lambda\to\infty,  
\end{equation}
we obtain from $N_{\lambda,\mu}
\left(
\theta_\star,\sigma,\theta_*\right)t^{|\lambda|+|\mu|}$
\begin{align*}
N_{\lambda,\mu}
\left(
\theta_\star,\sigma\right)t^{|\lambda|+|\mu|}
=&t^{|\lambda|+|\mu|}\prod_{(i,j)\in\lambda}
\frac{(\theta_\star+\sigma+i-j)}
{h_\lambda^2(i,j)(\lambda_j'+\mu_i-i-j+1+2\sigma)^2}
\\
&\times\prod_{(i,j)\in\mu}
\frac{(\theta_\star-\sigma+i-j)}
{h_\mu^2(i,j)(\mu_j'+\lambda_i-i-j+1-2\sigma)^2}. 
\end{align*}
 The degenerate function 
$\sum_{\lambda, \mu\in \mathbb{Y}}N_{\lambda,\mu}
\left(
\theta_\star,\sigma\right)t^{|\lambda|+|\mu|}$ 
is the builiding block of the tau function of the third Painlev\'e function 
of type $D_7^{(1)}$ at $t=0$ \cite{GIL2}. 

Finally,  by the limit
\begin{equation}\label{(1/2)1to(1/2)(1/2)}
\theta_\star=\Lambda,\quad t\to \frac{t}{\Lambda},\quad \Lambda\to\infty,  
\end{equation}
we obtain from $N_{\lambda,\mu}
\left(
\theta_\star,\sigma\right)t^{|\lambda|+|\mu|}$
\begin{align*}
N_{\lambda,\mu}
\left(\sigma\right)t^{|\lambda|+|\mu|}
=&t^{|\lambda|+|\mu|}\prod_{(i,j)\in\lambda}
\frac{1}
{h_\lambda^2(i,j)(\lambda_j'+\mu_i-i-j+1+2\sigma)^2}
\\
&\times\prod_{(i,j)\in\mu}
\frac{1}
{h_\mu^2(i,j)(\mu_j'+\lambda_i-i-j+1-2\sigma)^2}. 
\end{align*}
 The degenerate function 
$\sum_{\lambda, \mu\in \mathbb{Y}}N_{\lambda,\mu}
\left(\sigma\right)t^{|\lambda|+|\mu|}$ 
is the builiding block of the tau function of the third Painlev\'e function 
of type $D_8^{(1)}$ at $t=0$ \cite{GIL2}.

The limiting procedures \eqref{0000to100}, \eqref{100to11}, 
\eqref{100to(1/2)00}, \eqref{11to(1/2)1} and \eqref{(1/2)1to(1/2)(1/2)} 
correspond to the following degeneration scheme
\begin{equation*}
\begin{diagram}
\node{(0,0,0,0)}\arrow{e}
\node{(1,0,0)}\arrow{e}\arrow{se}
\node{(1,1)}\arrow{e}
\node{(1/2,1)}\arrow{e}
\node{(1/2,1/2)}
\\
\node[3]{(1/2,0,0)}\arrow{ne}
\end{diagram}
\end{equation*}
where $0,1,1/2$ represent the ranks of irregularities. 

\subsection{Rearranged expansion}

So far, local expansions of irregular conformal blocks or 
degeneration limits of the Nekrasov partition functions 
have been obtained when their expansions are at regular singular points. 
In general, degeneration limits of series expansions at 
irregular singular points are more delicate. 
To our knowledge, expansions at irregular singular points 
had been only considered 
in \cite{GT} before \cite{N ICB}. 
Consider 
\begin{equation}
|R^{(2)}\rangle=\Phi_{\Delta_4,\Delta}^{\Delta_3}(w)\Phi_{\Delta,\Delta_1}^{\Delta_2}(z)|\Delta_1\rangle,
\end{equation}
where $\Delta_i=\alpha_i(Q-\alpha_i)$. In what follows, 
we let $w$ go to zero, while $z$ is in a general position. Then 
$|R^{(2)}\rangle$ becomes an expansion of $z$ at the irregular singular point zero. 

We already know how to take a limit of 
$\Phi_{\Delta_4,\Delta}^{\Delta_3}(w)$ (see Subsection 3.2 and 3.3) 
but in the limit \eqref{eq_L_1-L} and \eqref{eq_L_2-L}, the coefficients $R_k$ of $z^k$ in 
\begin{equation*}
|R^{(2)}\rangle=z^{\Delta-\Delta_2-\Delta_1}
w^{\Delta_4-\Delta_3-\Delta}\sum_{k=0}^\infty R_kz^k
\end{equation*}
diverge. 
Instead, 
Gaiotto and Teschner suggested a 
 rearranged expansion of $|R^{(2)}\rangle$:  
\begin{equation}
|R^{(2)}\rangle=z^{\Delta-\Delta_2-\Delta_1}
\left(1-\frac{z}{w}\right)^A
\sum_{k=0}^\infty z^k |R_k^{(1)}\rangle
\end{equation}
for some constant $A$ in Appendix D of \cite{GT}. Note that 
$|R_0^{(1)}\rangle=\Phi_{\Delta_0,\Delta}^{\Delta_1}(w)|\Delta\rangle$.  
The condition of the limit \eqref{eq_L_1-L} and \eqref{eq_L_2-L} is now
\begin{equation}\label{eq-limit-rank1-vector}
\Delta_3-\Delta=\frac{c_1}{\epsilon}+c_{10}+O(\epsilon),
\quad
2\Delta_3-\Delta=\frac{c_2}{\epsilon^2}+\frac{c_{21}}{\epsilon}+c_{20}+O(\epsilon)\quad (\epsilon\to 0). 
\end{equation}
The resulting vector 
$|I^{(1)}\rangle=\lim_{\epsilon\to0}
w^{-\Delta_4+\Delta_3+\Delta}|R_0^{(1)}\rangle$ as 
$w= \epsilon$
satisfies
\begin{equation*}
L_1|I^{(1)}\rangle=c_1|\Lambda\rangle,\quad
L_2|I^{(1)}\rangle=c_2|\Lambda\rangle. 
\end{equation*}
Also $ |R_k^{(1)}\rangle$
satisfy 
\begin{align}
(L_0-w\partial_w-\Delta_3-\Delta)|R_k^{(1)}\rangle
=&k|R_k^{(1)}\rangle,\nonumber
\\
(L_n-w^n(w\partial_w+(n+1)\Delta_3))|R_k^{(1)}\rangle
=&A\sum_{s=1}^{n-1}w^{n-s}|R_{k-s}^{(1)}\rangle \label{eq_L_n_prerank1}
\\
&+(A+\Delta+n\Delta_2-\Delta_1+k-n)|R_{k-n}^{(1)}\rangle
\quad (n>0), \nonumber
\end{align}
which are derived using the commutation relations 
\eqref{comrel_rank0}. 
It is easy to see that the coefficients of 
vectors in the right hand side of 
the recursion relations admit a limit by
\begin{equation}\label{eq_limit_A}
A=O(\epsilon^{-1}),\quad A+\Delta-\Delta_1=O(1),\quad \epsilon\to 0. 
\end{equation}
By rescaling $|R_k^{(1)}\rangle$ as 
$|\widetilde{R}_k^{(1)}\rangle=w^{-\Delta_0+\Delta_1+\Delta}|R_k^{(1)}\rangle$, 
the coefficient in the left hand side of 
the recursion relations admits a limit  
 by 
 \eqref{eq-limit-rank1-vector}. 
Hence, together with \eqref{eq-limit-rank1-vector} and 
\eqref{eq_limit_A}, 
we can take a limit of the above 
 recursion relations. 
 
 Now let us see how we can take a limit 
of $|\widetilde{R}_1^{(1)}\rangle$. Comparing 
the normal expansion and the rearranged expansion 
of $|R^{(2)}\rangle$, we obtain 
\begin{equation}
|\widetilde{R}_1^{(1)}\rangle=
b_1L_{-1}|\widetilde{R}_0^{(1)}\rangle
-\frac{b_1}{w}L_0|\widetilde{R}_0^{(1)}\rangle
+\left(\frac{b_1}{w}(\Delta+\Delta_3)+\frac{A}{w}
\right)|\widetilde{R}_0^{(1)}\rangle,
\end{equation}
where 
\begin{equation*}
b_1=\frac{\Delta+\Delta_2-\Delta_1}{2\Delta}. 
\end{equation*}
It is easy to see that $b_1$ and $b_1/w$ have a finite limit. 
The final term has a finite limit if and only if 
\begin{equation}\label{eq_limit_A_2}
A=-\frac{\beta_1}{\epsilon}+O(1),\quad 
A+\Delta-\Delta_1=-\frac{c_1\beta_1}{2c_2}-\Delta_2+O(\epsilon)
,\quad \epsilon\to 0. 
\end{equation} 
We have observed  that $|\widetilde{R}_k^{(1)}\rangle$ ($k>1$) 
converge without additional conditions. 
 We remark that the limit of 
 the recursion relation \eqref{eq_L_n_prerank1} 
 under \eqref{eq-limit-rank1-vector}  and \eqref{eq_limit_A_2} 
is  equal to \eqref{eq_L_n_rank1}. 
 
 The remained task is to prove that $|\widetilde{R}_k^{(1)}\rangle$ 
 has a finite limit. In \cite{GT},  it was claimed that 
 uniqueness of elements satisfying the all resulting recursion relations 
\eqref{eq_L_n_rank1}  yields convergence of $|\widetilde{R}_k^{(1)}\rangle$. 
 Although the uniqueness was proved in Theorem \ref{thm_VO_rr} (\cite{N ICB}), 
we could not confirm their statement. We might need  another approach and 
hope to report it in the near future. 
 
 \section{Direct approach}

In the previous studies, by degeneration limits from regular 
conformal blocks, irregular conformal blocks 
have been obtained as local expansions at regular singular points 
or integral representations \cite{Nishinaka Rim 2012}, \cite{Choi Rim 2015}. There was also an attempt to 
obtain expansions of irregular conformal blocks at 
irregular singular points by collision limits \cite{GT}. However, 
it should be better to have a direct method to obtain  
irregular conformal blocks, neither by confluence process nor by 
asymptotic expansions of integrals. 
 
In what follows, we present a definition of 
irregular versions of vertex operators and define 
irregular conformal blocks directly, which were proposed in \cite{N ICB}. 

For $r\in\Z_{\ge 0}$, denote the Whittaker module by $W^{[r]}_\Lambda$ such that  
\begin{align*}
&L_n|\Lambda\rangle=\Lambda_n|\Lambda\rangle\quad (n=r,r+1,\ldots, 2r),
\end{align*}
with $\Lambda=(\Lambda_r,\ldots, \Lambda_{2r})$  
and $W^{[r]}_\Lambda$ is spanned by 
 linearly independent vectors 
  of the form
\begin{equation*}
L_{i_1}\cdots L_{i_k}|\Lambda \rangle \quad (i_1\le \cdots\le i_k<r). 
\end{equation*}  
Denote the dual Whittaker module by $W^{*, [r]}_\Lambda$ such that
\begin{align*}
&\langle \Lambda|L_n=\Lambda_n\langle\Lambda|\quad (n=-r,-r-1,\ldots, -2r),
\end{align*}
with $\Lambda=(\Lambda_{-r},\ldots, \Lambda_{-2r})$ 
and $W^{*,[r]}_\Lambda$ is spanned by linearly independent 
 vectors 
  of the form
\begin{equation*}
\langle\Lambda|L_{i_1}\cdots L_{i_k}\quad (-r<i_1\le \cdots\le i_k). 
\end{equation*}
See, for example, \cite{FJK}, \cite{LGZ} on the details 
of Whittaker modules. 

A bilinear pairing $\langle | \rangle$: $ W^{*,[0]}_{\Lambda'}\times W^{[1]}_\Lambda\to \C$ is 
uniquely defined by 
\begin{align*}
&\langle \Lambda'| \cdot |\Lambda\rangle=1,
\\
&\langle u|L_n \cdot |v\rangle =\langle u| \cdot L_n |v\rangle \equiv 
\langle u| L_n |v\rangle, 
\end{align*}
where $\langle u|\in W^{*,[0]}_{\Lambda'}$, $|v\rangle\in W^{[1]}_\Lambda$. 
Because, if $n>0$, then $L_n$ acts on $|\Lambda\rangle$ diagonally and if $n\le 0$, then 
$L_n$  acts on $\langle \Lambda'|$ diagonally. Let $V^{*,[0]}_0$ be the irreducible highest weight 
representation. Then, 
a bilinear pairing $\langle |\rangle$: $ V^{*,[0]}_0\times W^{[2]}_\Lambda\to \C$ is 
also uniquely defined by 
\begin{align*}
&\langle 0| \cdot |\Lambda\rangle=1,
\\
&\langle u|L_n \cdot |v\rangle =\langle u| \cdot L_n |v\rangle \equiv 
\langle u| L_n |v\rangle, 
\end{align*}
where $\langle u|\in V^{*,[0]}_0$, $|v\rangle\in W^{[2]}_\Lambda$ 
because $\langle 0|L_1=0$. For higher cases, one way of defining a pairing 
is to use the Fock space, where a pairing is defined naturally. 

\begin{dfn}
For positive integer $r$, we define a vertex operator 
$\Phi^{\Delta}_{\Lambda',\Lambda}(z): 
W^{[r]}_{\Lambda}\to W^{[r]}_{\Lambda'}$  by 
\begin{align}
&[L_n, \Phi^{ \Delta}_{\Lambda',\Lambda}(z)]=z^n \left(z\frac{\partial}{\partial z}+
(n+1)\Delta\right)\Phi^{ \Delta}_{\Lambda',\Lambda}(z), 
\label{eq_com_VO_rr}
\\
&\Phi^{\Delta}_{\Lambda',\Lambda}(z)\Lam=z^\alpha \exp\left(\sum_{n=0}^r 
\frac{\beta_n}{z^n}\right)\sum_{n=0}^\infty
w_nz^n, \label{eq_act_VO_rr}
\end{align}
where $\alpha,\beta_n\in\C$, $w_n\in W^{[r]}_{\Lambda'}$ and 
$w_0=|\Lambda'\rangle$. 
\end{dfn}

For the case of $r=1$, by the commutation relation 
\eqref{eq_com_VO_rr} and condition \eqref{eq_act_VO_rr}, 
$w_n$ should satisfy 
\begin{equation}\label{eq_L_n_rank1}
(L_n-\delta_{n,1}\Lambda_1-\delta_{n,2}\Lambda_2)w_m
=-\beta_1w_{m+1-n}+(\alpha+(n+1)\Delta+m-n)w_{m-n}
\end{equation}
and
\begin{equation*}
\alpha=-\frac{\beta_1(\Lambda_1-\beta_1)}
{2\Lambda_2}-2\Delta_2,\quad \Lambda_1'=\Lambda-\beta_1,\quad 
\Lambda_2'=\Lambda_2. 
\end{equation*} 

For a partition 
$\lambda=(\lambda_1,\ldots,\lambda_n)$ ($\lambda_i\ge \lambda_{i+1}$), define  
$L_{-\lambda}=L_{-\lambda_1+r}\cdots
L_{-\lambda_n+r}$. By PBW theorem,  the set consisting of the vectors 
$L_{-\lambda}|\Lambda\rangle$ where $\lambda$ 
runs over all partitions, 
is a basis of $W^{[r]}_{\Lambda}$. Let 
$U_m$ be the subspace generated by $L_{-\lambda}|\Lambda\rangle$ 
such that $|\lambda|\le m$.

  \begin{thm}[\cite{N ICB}]\label{thm_VO_rr}
  For any positive integer $r$ and non-zero $\Lambda_{2r}$, 
the rank $0$ vertex operator $\Phi^{ \Delta}_{\Lambda',\Lambda}(z)$: 
$W^{[r]}_{\Lambda}\to W^{[r]}_{\Lambda'}$ exists 
and is  uniquely determined by the given 
parameters $\Lambda$, $\Delta$, $\beta_r$. 
In particular, 
\begin{equation*}
\Lambda'_n=\Lambda_n-\delta_{n,r}r\beta_r \quad (n=r,\ldots,2r),  
\end{equation*}
$\alpha$ and $\beta_n$ for $n=1,\ldots, r-1$ 
are  polynomials 
in $ \Delta$, $\beta_r$, $\Lambda_r$, 
\ldots, $\Lambda_{2r}$, $\Lambda_{2r}^{-1}$.  
Moreover,  $v_m\in U_m$ and 
the coefficients $c_\lambda$ of the vectors $
L_{-\lambda}|\Lambda'\rangle$ in $v_m$ are 
uniquely determined as polynomials 
in $ \Delta$, $\beta_r$, $\Lambda_r$,\ldots, $\Lambda_{2r}$, 
$\Lambda_{2r}^{-1}$.  
\end{thm}

Note that by scaling the variable $z$, we can remove $\Lambda_{2r}^{-1}$ 
in the coefficients. 
As a result, the coefficients of $\Phi^{\Delta}_{\Lambda',\Lambda}(z)\Lam$ are 
polynomials in $ \Delta$, $\beta_r$, $\Lambda_r$, 
\ldots, $\Lambda_{2r}$. In contrast,  
the coefficients of 
$\Phi^{\Delta_2}_{\Delta_3,\Delta_1}(z)|\Delta_1\rangle$ are 
rational functions in $\Delta_3$.  

Irregular conformal blocks are defined as expectation values 
of irregular vertex operators in the same way of regular conformal blocks. 
A three-point irregular conformal block with two 
regular singular points $z$, $\infty$ and one irregular singular point $0$ of rank $1$ 
is defined by 
\begin{equation}\label{PV regular}
\left( \langle \Delta | \Phi^{*,\Delta_2}_{\Delta,\Delta_1}(z)\right)\cdot  |\Lambda\rangle, 
\end{equation}
or  
\begin{equation}\label{PV irregular}
\langle \Delta| \cdot \left(\Phi^{\Delta_2}_{\Lambda', \Lambda}(z) |\Lambda\rangle\right), 
\end{equation}
where $\Phi^{*,\Delta_2}_{\Delta,\Delta_1}(z)$: $V^*_\Delta\to V^*_{\Delta_1}$ 
is the dual vertex operator and $\Lambda=(\Lambda_1,\Lambda_2)$ ($\Lambda_2\neq 0$). 
Although for the regular four-point case, the following two conformal blocks 
\begin{equation*}
\left( \langle \Delta_4 | \Phi^{*,\Delta_3}_{\Delta_4,\Delta}(z_3)\right)\cdot 
\left( \Phi^{\Delta_2}_{\Delta,\Delta_1}(z_2)|\Delta_1\rangle\right),\quad 
 \langle \Delta_4 |\cdot \left( 
 \Phi^{\Delta_3}_{\Delta_4,\Delta}(z_3)|\Delta_1\rangle
 \circ
 \Phi^{\Delta_2}_{\Delta,\Delta_1}(z_2)|\Delta_1\rangle
 \right)
\end{equation*}
are equal, the three-point irregular conformal blocks \eqref{PV regular}, 
\eqref{PV irregular} are different.  The former is an expansion at 
the regular singular point $\infty$ and the latter is an expansion at 
the irregular singular point $0$. 

A two-point irregular conformal block with 
one regular singular point $z$ and one irregular singular point $0$ of rank $2$ 
is defined by 
\begin{equation}\label{PIV irregular}
\langle 0| \cdot \left( \Phi^{\Delta_2}_{\Lambda',\Lambda}(z) |\Lambda\rangle\right), 
\end{equation}
where $\Lambda=(\Lambda_2,\Lambda_3,\Lambda_4)$ ($\Lambda_4\neq 0$). 

\section{Conjectures }
\subsection{Series expansion formulas of $\tau_{\mathrm{V}}(t)$ and 
$\tau_{\mathrm{IV}}(t)$}
In this section, we assume the central charge $c=1$. 
As the tau function of $\mathrm{P_{VI}}$ 
is expressed as a Fourier expansion \eqref{PVI regular} in terms of a four-point 
regular conformal block, we expect that expansions at 
the irregular singular point $\infty$ of 
the tau functions of $\mathrm{P_{V}}$, $\mathrm{P_{IV}}$  
are  Fourier expansions in terms of a three-point 
irregular conformal block with two 
regular singular points $z$, $\infty$ 
and one irregular singular point $0$ of rank $1$ \eqref{PV irregular}, 
a two-point irregular conformal block with 
one regular singular point $z$ and one irregular singular point $0$ of rank $2$ 
\eqref{PIV irregular}, respectively. 

Recall that the Hamiltonian functions $H_\mathrm{J}(t)$ of $\mathrm{P_J}$ 
($J=\mathrm{V}, \mathrm{IV}$)   
satisfy  
\begin{align}
&(th_\mathrm{V}'')^2-(h_\mathrm{V}-th_\mathrm{V}'+2(h_\mathrm{V}')^2)^2+\frac{1}{4}((2h_\mathrm{V}'-\theta)^2-4\theta_0^2)
((2h_\mathrm{V}'+\theta)^2-4\theta_t^2)=0,
\label{eq-Hamiltonian-DE-V} 
\\
&\left(H_\mathrm{IV}''\right)^2-4(tH_\mathrm{IV}'-H_\mathrm{IV})^2+4H_\mathrm{IV}'(H_\mathrm{IV}'-2(\theta+\theta_t))(H_\mathrm{IV}'
-4\theta_t)=0, 
\label{eq-Hamiltonian-DE-IV}
\end{align}
where $f'=df/dt$ and 
$h_\mathrm{V}=tH_\mathrm{V}$. The $\tau$-functions $\tau_\mathrm{J}=\tau_\mathrm{J}(t)$ are related to the Hamiltonian functions as 
\begin{equation*}
H_\mathrm{IV}(t)=\frac{d}{dt}\log \tau_\mathrm{IV}(t),\quad 
H_\mathrm{V}(t)=t\frac{d}{dt}\log \tau_\mathrm{V}(t).  
\end{equation*} 
Note that the tau functions satisfying these differential equations 
define each Painlev\'e functions. 

Based on the expectations, we substitute 
\begin{equation*}
\sum_{n\in\mathbb{Z}}s^n C_n \mathcal{F}(\theta, \Lambda, \beta_n;t),
\end{equation*} 
where 
\begin{equation*}
\mathcal{F}(\theta,\Lambda, \beta_n;t)=
\langle\theta_0^2|
\cdot \left( \Phi^{\theta_t^2}_{(\Lambda_1-\beta_n,\Lambda_2), 
(\Lambda_1,\Lambda_2)}
(t^{-1})|(\Lambda_1, \Lambda_2)\rangle\right)
\end{equation*}
for $\mathrm{P_V}$ case, into \eqref{eq-Hamiltonian-DE-V} and 
\begin{equation*}
\langle 0|\cdot \left( \Phi^{\theta_t^2}_{(\Lambda_2-\beta_n,\Lambda_3,\Lambda_4),(\Lambda_2, \Lambda_3,\Lambda_4)}(t^{-1})
|(\Lambda_2, \Lambda_3,\Lambda_4)\rangle\right)
\end{equation*}
for $\mathrm{P_{IV}}$ case, into \eqref{eq-Hamiltonian-DE-IV}. 
Then, we look at the coefficient of $s^i$ for $i\in\Z$ which is of the form 
\begin{equation*}
 t^A e^B(a_0+a_1t^{-1}+a_2t^{-2}+\cdots).
 \end{equation*}
Since $a_i$ ($i=0,1,2,\ldots,$) are polynomials in the parameters, 
we can determine $C_n$, $\beta_n$, $\Lambda_i$ in terms of 
$\theta_i$ and $\beta$. 

\begin{conj}[$\mathrm{P_{V}}$ case, \cite{N ICB}]
Let 
\begin{align*}
\tau(t)=\sum_{n\in\Z}&s^n(-1)^{n(n+1)/2}
G(1\pm\theta_0+\theta-\beta-n)G(1+\theta_t
\pm(\beta+n))\nonumber
\\
&\times \langle \theta_0^2|
\cdot \left( \Phi^{\theta_t^2}_{(\theta-\beta-n,1/4),(\theta,1/4)}
(t^{-1})|(\theta,1/4)\rangle\right)
\end{align*}
and $H=t (\log(t^{-2\theta_t^2-\theta^2/2}e^{-\theta t/2}\tau(t)))'$. Then, 
$H$ satisfies the differential equation
\eqref{eq-Hamiltonian-DE-V}. 
\end{conj}
\begin{conj}[$\mathrm{P_{IV}}$ case, \cite{N ICB}] 
Let
\begin{align*}
\tau(t)=t^{-2\theta_t^2}e^{\theta_t t^2}\sum_{n\in\Z}
s^n &G(1+\theta-\beta-n)
\prod_{\epsilon=\pm 1}
G(1+\theta_t+\epsilon(\beta+n))
\\
&\times \langle 0|\cdot \left( \Phi^{\theta_t^2}_{(\theta,0,1/4),(\theta-\beta-n, 0,1/4)}(1/\sqrt{2}t)
|(\theta,0,1/4)\rangle\right)
\end{align*}
and $ H=(\log \tau(t))'$. Then, $H$ satisfies the  
differential equation 
\eqref{eq-Hamiltonian-DE-IV}. 
\end{conj}
Here, 
\begin{align}
&\langle \theta_0^2|
\cdot \left( \Phi^{\theta_t^2}_{(\theta,1/4),(\theta-\beta,1/4)}
(1/t)|(\theta,1/4)\rangle\right)\label{ICB rank one}
\\
&=t^{2\theta_t^2+2\beta(\theta-\beta)}
e^{\beta t}\left(1+2 \left(2 \beta^3-3 \beta^2 \theta +\beta  \theta^2-\beta  \theta_0^2-\beta  \theta_t^2+\theta  \theta_t^2\right)t^{-1} \right.\nonumber
\\
&+2 \left(4 \beta ^6-12 \beta ^5 \theta +13 \beta ^4 \theta ^2-4 \beta ^4 \theta_0^2-4 \beta ^4 \theta_t^2+5 \beta ^4-6 \beta ^3 \theta ^3+6 \beta ^3 \theta  \theta_0^2+10 \beta ^3 \theta  \theta_t^2-10 \beta ^3 \theta +\beta ^2 \theta ^4
\right.\nonumber
\\
&-2 \beta ^2 \theta ^2 \theta_0^2-8 \beta ^2 \theta ^2 \theta_t^2+6 \beta ^2 \theta ^2+\beta ^2 \theta_0^4+2 \beta ^2 \theta_0^2 \theta_t^2-3 \beta ^2 \theta_0^2+\beta ^2 \theta_t^4-3 \beta ^2 \theta_t^2+2 \beta  \theta ^3 \theta_t^2-\beta  \theta ^3\nonumber
\\
&\left. \left.-2 \beta  \theta  \theta_0^2 \theta_t^2+\beta  \theta  \theta_0^2-2 \beta  \theta  \theta_t^4+5 \beta  \theta  \theta_t^2+\theta ^2 \theta_t^4-2 \theta ^2 \theta_t^2+\theta_0^2 \theta_t^2\right)t^{-2}+\cdots \right)\nonumber
\end{align}
and 
\begin{align*}
&\langle 0|\cdot \left( \Phi^{\theta_t^2}_{(\theta,0,1/4),(\theta-\beta, 0,1/4)}(1/t)
|(\theta,0,1/4)\rangle\right)
\\
&=t^{3\theta_t^2+\beta(2\theta-3\beta)}
e^{\beta t^2/2}\left(1+
\left(\theta ^2 \beta +2 \theta  \theta_t^2-6 \theta  \beta ^2-3 \theta_t^2 \beta +6 \beta ^3\right)t^{-2}
\right.
\\
&+\frac{1}{4} \left(2 \theta ^4 \beta ^2+8 \theta ^3 \theta_t^2 \beta -24 \theta ^3 \beta ^3-4 \theta^3 \beta +8 \theta ^2 \theta_t^4-60 
\theta^2 \theta_t^2 \beta^2-16 \theta^2 \theta_t^2+96 \theta^2 \beta^4\right. 
\\
&+48 \theta^2 \beta^2-24 \theta  \theta_t^4 \beta +120 \theta  \theta_t^2 \beta^3+72 \theta  \theta_t^2 \beta -144 \theta  \beta^5-140 \theta  \beta^3-2 \theta  \beta +18 \theta_t^4 \beta ^2
\\
&\left.\left.+\theta_t^4-72 \theta_t^2 \beta^4-66 \theta_t^2 \beta^2-\theta_t^2+72 \beta^6+105 \beta^4+3 \beta^2\right)t^{-4}+\cdots \right). 
\end{align*}

\subsection{A combinatorial expansion formula 3-pt ICB at  an irregular singular point}
It is natural to expect that these irregular conformal blocks have combinatorial 
expressions, as expansions of conformal blocks at regular singular points have 
explicit formulas parametrized by partitions. To our knowledge, in general case, 
such combinatorial 
formulas for expansions of irregular conformal blocks at irregular singular points had  
not been reported. 

In what follows, we only consider 
the irregular conformal block of rank one for $\mathrm{P_V}$. Note that 
a particular three-point irregular conformal block of rank one  
is the confluent hypergeometric function, whose coefficients of $t^{-k}$ are 
factorized. Let us substitute $\beta$ into $\theta_t$ in \eqref{ICB rank one}. Then, 
the coefficient of $t^{-1}$ becomes
$2 \beta \left((\theta -\beta )^2-\theta_0^2\right)$. If we substitute 
$\theta-\beta$ 
into $\theta_0$, we see that the coefficient of $t^{-1}$ becomes 
$2 (\beta-\theta ) \left(\beta ^2-\theta_t^2\right)$. Fortunately, we have 
\begin{equation*}
2 \left(2 \beta^3-3 \beta^2 \theta +\beta  \theta^2-\beta  \theta_0^2-\beta  \theta_t^2+\theta  \theta_t^2\right)=2 (\beta-\theta ) \left(\beta ^2-\theta_t^2\right)+2 \beta  \left((\theta -\beta )^2-\theta_0^2\right). 
\end{equation*}
Thus, the coefficient of $t^{-1}$ is successfully written as a sum of two factorized forms, 
which should be corresponding to pairs of partitions
($(1)$, $\emptyset$), ($\emptyset$, $(1)$). These two factors look like 
same as the factors in the partition functions in Section 3.3. We can guess that 
a pair ($\lambda$, $\emptyset$) of partitions is corresponding to 
\begin{equation*}
N_{\lambda,\emptyset}=\prod_{(i,j)\in\lambda}
\frac{\left(2(\beta-\theta)+i-j\right)\left((\beta+i-j)^2-\theta_t^2\right)}
{h_\lambda(i,j)^2}
\end{equation*}
and 
a pair ($\emptyset$, $\mu$) of partitions is corresponding to 
\begin{equation*}
N_{\emptyset,\mu}=(-1)^{|\mu| }\prod_{(i,j)\in\mu}
\frac{\left(-2\beta+i-j\right)\left((\theta-\beta+i-j)^2-\theta_0^2\right)}
{h_\mu(i,j)^2}. 
\end{equation*}
Let us check that this assumption works for the coefficient of $t^{-2}$ 
which should be expressed by a sum of five factors corresponding to 
pairs of partitions ($(2),\emptyset$), ($(1,1),\emptyset$), ($(1),(1)$), 
($\emptyset,(2)$), ($\emptyset, (1,1)$). We see that the coefficient of $t^{-2}$ 
is equal to 
\begin{align*}
&\sum_{|\lambda|=2}
\left(N_{\lambda,\emptyset}
+N_{\emptyset,\lambda}\right)
+2(2(\theta-\beta)\beta-1)\left(\beta ^2-\theta_t^2\right) \left((\theta -\beta )^2-\theta_0^2\right).  
\end{align*}
The last term should be corresponding to ($(1),(1)$) but 
contains a quadratic form which is not factored by linear equations. 
If we use $N_{(1),(1)}$, then the last term is expressed as 
\begin{equation*}
N_{(1),(1)}-2\left(\beta ^2-\theta_t^2\right) \left((\theta -\beta )^2-\theta_0^2\right). 
\end{equation*}
Hence, the term corresponding to ($(1), (1)$) 
is written by a sum of two factored forms. Keeping this in mind, we put 
\begin{align*}
&U_\lambda=\prod_{(i,j)\in\lambda}
(2(\beta-\theta)+i-j),\quad V_\lambda=\prod_{(i,j)\in\lambda}
(-2\beta+i-j),
\\
&S_{\lambda,\mu}=(-1)^{|\mu|}
\prod_{(i,j)\in\lambda}\frac{(\beta+i-j)^2-\theta_t^2}{h_\lambda(i,j)^2}
\prod_{(i,j)\in\mu}\frac{(\theta-\beta+i-j)^2-\theta_0^2}{h_\mu(i,j)^2}.   
\end{align*}
Then, 
 the coefficient of $t^{-3}$ is equal to 
\begin{align*}
&\sum_{|\lambda|=3}
\left(N_{\lambda,\emptyset}
+N_{\emptyset,\lambda}\right)
+\sum_{|\lambda|=2}\left(
N_{\lambda,(1)}-4U_{\lambda/(1)}S_{\lambda,(1)}
+N_{(1),\lambda}-4V_{\lambda/(1)}S_{(1),\lambda}
\right), 
\end{align*}
where, $U_{\lambda/\mu}=U_{\lambda}/U_{\mu}$ and 
$V_{\lambda/\mu}=V_{\lambda}/V_{\mu}$. 
Similarly, 
the coefficient of $t^{-4}$ is expressed as 
\begin{align*}
&\sum_{|\lambda|=4}
\left(N_{\lambda,\emptyset}
+N_{\emptyset,\lambda}\right)
+\sum_{|\lambda|=3}
\left(N_{\lambda,(1)}
-6U_{\lambda/(1)}S_{\lambda,(1)}
+N_{(1),\lambda}
-6V_{\lambda/(1)}S_{(1),\lambda}\right)
\\
&+N_{(2),(2)}-8U_{(2)/(1)}V_{(2)/(1)}S_{(2),(2)}+4S_{(2),(2)}
\\
&
+N_{(2),(1,1)}-8U_{(2)/(1)}V_{(1,1)/(1)}S_{(2),(1,1)}+12S_{(2),(1,1)}
\\
&
+N_{(1,1),(2)}-8U_{(1,1)/(1)}V_{(2)/(1)}S_{(1,1),(2)}+12S_{(1,1),(2)}
\\
&
+N_{(1,1),(1,1)}-8U_{(1,1)/(1)}V_{(1,1)/(1)}S_{(1,1),(1,1)}+4S_{(1,1),(1,1)}. 
\end{align*}

Note that $N_{\lambda,\mu}=U_{\lambda/\emptyset}V_{\mu/\emptyset}S_{\lambda,\mu}$. 
Based on the observations, we propose the next conjecture. 
\begin{conj}
A three-point irregular conformal block with 
two regular singular points $z$, $\infty$ and one irregular singular point 
$0$ of rank one admits the following combinatorial formula
\begin{align*}
&\langle \theta_0^2|
\cdot \left( \Phi^{\theta_t^2}_{(\theta,1/4),(\theta-\beta,1/4)}
(t)|(\theta,1/4)\rangle\right)
\\
&=t^{-2\theta_t^2-2\beta(\theta-\beta)}
e^{\frac{\beta}{t}}\sum_{\lambda,\mu\in\mathbb{Y}}t^{|\lambda|+|\mu|}
\sum_{\nu\subset\lambda,\eta\subset\mu,\atop |\nu|=|\eta|}
(-1)^{|\nu|}
c_{\lambda,\mu}^{\nu,\eta}
U_{\lambda/\nu}V_{\mu/\eta}S_{\lambda,\mu},
\end{align*}
where $c_{\lambda,\mu}^{\nu,\eta}\in\Z_{\ge 0}$, 
as an expansion at the irregular singular point $0$. 
\end{conj}

As long as we calculate, the coefficients $c_{\lambda,\mu}^{\nu,\eta}$ 
are non-negative integers. Further, we observe 
\begin{align*}
&c_{\lambda,\mu}^{\emptyset,\emptyset}=1,
\quad c_{\lambda,\mu}^{(1),(1)}=2|\lambda||\mu|,
\quad c_{\lambda,\mu}^{(2),(2)}=q_\lambda q_\mu,\quad 
c_{\lambda,\mu}^{(2),(1,1)}=3q_\lambda q_{\mu'},
\quad c_{\lambda,\mu}^{\nu,\eta}=c_{\mu,\lambda}^{\eta,\nu}
=c_{\lambda',\mu'}^{\nu',\eta'},
\end{align*}
where 
\begin{equation*}
q_\lambda=\lambda_1(\lambda_1-1)
+\sum_{(i,j)\in\lambda,\atop i\neq 1}\left(\lambda_1-1+\sum_{k=1}^{j-1}\lambda'_k
\right). 
\end{equation*}
The function $q_\lambda$ of partitions can be regarded as  
a function of particular (column) semi-standard tableaux such that 
for $\lambda$, ($1,j$) box has $2(j-1)$ and ($i,j$) box for $i\neq 1$ 
has  $\lambda_1-1+\sum_{k=1}^{j-1}\lambda'_k$. 
As an example, for $\lambda=(6,4,4,3,2,1)$, the corresponding 
tableau is 
\begin{equation*}
\ytableausetup{centertableaux}
\begin{ytableau}
0 & 2 & 4  & 6 & 8 & 10 \\
5 & 11 & 16 & 20\\
5 & 11 & 16 & 20\\
5 & 11 & 16 \\
5 & 11 \\
5
\end{ytableau}. 
\end{equation*}

\end{document}